\documentclass[aps,pre,twocolumn,amsmath,amssymb,superscriptaddress,showpacs,longbibliography]{revtex4-1}

\usepackage{graphicx}
\usepackage{dcolumn}
\usepackage{bm}
\usepackage{mathrsfs}

\usepackage{natbib}
\usepackage[text={7in,9.5in},centering]{geometry}

\usepackage{hyperref}
\usepackage{url}

\usepackage{epsfig}
\usepackage{amsmath}
\usepackage{amssymb}
\usepackage{textcomp}

\usepackage{color}
\usepackage{float}

\usepackage{stackengine}
\usepackage{verbatim}

\begin{document}

\title{A new anomaly observed in $^4$He supports the existence of 
the hypothetical X17 particle}

\author{A.J. Krasznahorkay}
\email{kraszna@atomki.hu}
\author{M. Csatl\'os}
\author{L. Csige}
\author{J. Guly\'as}
\author{A.~Krasznahorkay}
\altaffiliation{Currently working at CERN, Geneva, Switzerland}
\author{B.M. Nyak\'o}
\author{I. Rajta}
\author{J. Tim\'ar}
\author{I. Vajda}
\affiliation{Institute of Nuclear Research  (ATOMKI),
  P.O. Box 51, H-4001 Debrecen, Hungary} 
\author{N.J. Sas}
\affiliation{University of Debrecen, 4010 Debrecen, PO Box 105, Hungary}

\begin{abstract}

Energy-sum and angular correlation spectra of $e^+e^-$ pairs produced
in the $^{3}$H(p,$\gamma$)$^{4}$He nuclear reaction have been studied
at $E_p$=510, 610 and 900 keV proton energies.  The main features
of the spectra can be understood by taking into account the internal
and external pair creations following the direct proton radiative
capture by $^{3}$H.  However, these processes cannot                                
account for the observed peak around 115$^\circ$ in the angular
correlation spectra. This anomalous excess of $e^+e^-$ pairs can be
described by the creation and subsequent decay of a light particle
during the direct capture process.  The derived mass of the particle
is $m_\mathrm{X}c^2$=16.94$\pm0.12 (stat) \pm 0.21 (syst)$~MeV.                     
According to the mass and branching ratio ($B_x=5.1(13)\times10^{-6}$),
this is likely the same X17 particle, which we recently suggested
[Phys. Rev. Lett. 116, 052501 (2016)] for describing the anomaly
observed in the decay of $^8$Be.

\end{abstract}


\maketitle

Recently, we measured electron-positron angular correlations for the
17.6 MeV, and 18.15 MeV, J$^p = 1^+ \rightarrow J^\pi = 0^+$, M1
transitions in $^8$Be and an anomalous angular correlation was
observed \cite{kr16}.  This was interpreted as the creation and decay
of an intermediate bosonic particle, which we now call X17, with a mass of          
$m_\mathrm{X}c^2$=16.70$\pm 0.35 $(stat) $\pm 0.5 $(sys) MeV.  The
possible relation of the X17 boson to the dark matter problem and the
fact that it might explain the (g-2)$_\mu$ puzzle, triggered an
enhanced theoretical and experimental interest in the particle and
hadron physics community \cite{da19,ins}. 
A number of such light particles have already been predicted for many 
decades with a wide range of different properties 
\cite{we78,wi78,do78,fa04,po08,wo10,da12}, but have never been 
confirmed experimentally.

Our data were first explained with a 16.7 MeV, vector gauge boson, X17               
by Feng and co-workers \cite{fe16,fe17}, which may mediate a fifth
fundamental force with some coupling to Standard Model (SM) particles.
The X17 boson is thus produced in the decay of an excited state to the
ground state, $^8$Be$^*\rightarrow^8$Be + X17, and then decays through
the X17 $\rightarrow$ $e^+e^-$ process. Constraints on such a new
particle, were also taken into account by Feng and co-workers
\cite{fe16,fe17}.

Zhang and Miller \cite{zh17} investigated the possibility to explain
the anomaly within nuclear physics. They explored the nuclear
transition form factor as a possible origin of the anomaly, and found
the required form factor to be unrealistic for the $^8$Be nucleus.

Ellwanger and Moretti made another possible explanation of the
experimental results through a light pseudo-scalar particle
\cite{ell16}. Given the quantum numbers of the $^8$Be$^*$ and $^8$Be
states, the X17 boson could indeed be a $J^\pi = 0^-$ pseudo-scalar
particle, if it was emitted with L = 1 orbital angular momentum. 

Alves and Weiner \cite{al18} revisited experimental constraints on QCD
axions in the O(10 MeV) mass window.  In particular, they found a
variant axion model that remains compatible with existing constraints.
This reopens the possibility of solving the strong CP problem at the
GeV scale.  Such axions or axion-like particles (ALPs) are expected to
decay predominantly by the emission of $e^+e^-$ pairs.

Subsequently, many studies with different models have been performed
including an extended two Higgs doublet model \cite{de17}.  Delle Rose
and co-workers \cite{de19} showed that the anomaly can be described
with a very light Z$_0$ bosonic state.  They also showed
\cite{ro19} how both spin-0 and spin-1 solutions are possible and
describe Beyond the Standard Model (BSM) scenarios.

In parallel to these recent theoretical studies, we re-investigated
the $^8$Be anomaly with an improved experimental setup. We have
confirmed the signal of the assumed X17 particle and measured its mass
[$m_\mathrm{X}c^2 = 17.01(16)$ MeV] and branching ratio compared to
the $\gamma$ decay [$B_x = 6(1)\times 10^{-6}$] with improved precision
\cite{kra17,kra19}. 
The observed deviation in the angular correlation of the $e^+e^-$ pairs 
was found much smaller in the 17.6 MeV than in the 18.15 MeV transition 
of $^8$Be \cite{kr17}.

In the present work, we have conducted a search for the X17 particle
in the $^{3}$H(p,$\gamma$)$^{4}$He reaction using different
proton beam energies.  The experiment was performed in Debrecen at the
2 MV Tandetron accelerator of ATOMKI. The $^{3}$H(p,$\gamma$)$^{4}$He
reaction was used at  bombarding energy of E$_p$ = 510, 610 and 900
keV to induce  radiative capture and to populate the            
overlapping $J^\pi=0^+$ first, and $J^\pi=0^-$ second excited states in $^4$He
\cite{ti18}.

The proton beam with a typical current of 1.0 $\mu$A was impinged on a
$^3$H target for about 100 hours for each bombarding energy.  The
$^3$H was absorbed in a 4.2~mg/cm$^2$ thick Ti layer evaporated onto a
0.4 mm thick molybdenum disc with a diameter of 50 mm. 
The density of the $^3$H atoms was $\approx 2.7\times10^{20}$ atoms/cm$^2$. 
The disk was cooled down to liquid
N$_2$ temperature to prevent $^3$H evaporation. In such a thick
$^3$H+Ti target, the proton beam was stopped completely. 
The target was shifted off the center of the spectrometer 
by 20 mm upstream to avoid the screening by the target backing and holder.
 We have used also a large (Duth) fixing screw, 
so the maximal correlation angle we could utilize was $\approx 140^o$. 

The proton energies were chosen to stay below the threshold of the
(p,n) reaction (E$_{thr}$ = 1.018 MeV). The $^4$He nucleus was excited
up to E$_x$ = 20.21, 20.29 and 20.49 MeV at the used proton beam energies,
so it was expected that the first excited state of $^4$He
(J$^p$=0$^+$, E$_x$=20.21~MeV, $\Gamma$=0.50 MeV) and the second one
(J$^p$=0$^-$, E$_x$=21.01~MeV, $\Gamma$=0.84 MeV) were also
populated in $^4$He \cite{ti18}.

Our previous experimental setup \cite{kr16,gu16} has recently been
upgraded by the replacement of the scintillators (EJ200) and 
PM tubes (Hamamatsu 10233-100) in order
to avoid the effects of scintillator material aging. As another
improvement, the MultiWire Proportional Chambers (MWPC)  have
been replaced by novel Double-sided Silicon Strip  Detectors  (DSSD)to
enhance the efficiency of the experimental setup and its homogeneity.

We also increased the
number of telescopes from 5 to 6. The sizes of the scintillators were
$82\times86\times 80$ mm$^3$ each.  The positions of the hits were
registered by the DSSDs having sizes of 50x50 mm$^2$ strip widths of 3
mm and a thickness of 500 $\mu$m.  The telescope detectors were placed
perpendicularly to the beam direction, each at 60$^\circ$ to its
neighbors, around a vacuum chamber made of a carbon fiber tube with a
wall thickness of 1 mm.

In order to search for the assumed X17 particle, both the energy-sum
spectrum of the $e^+e^-$ pairs measured by the telescopes, and their
angular correlations, determined by the DSSD detectors, have been
analyzed. For the real ``signal'' events we always requred that the 
energy-sum for the $e^+e^-$-pairs should be equal to the thansition energy,
which we want to investigate.  

 Since the counting rates in the detectors were low
($\approx 150$ Hz in the scintillators and $\approx 25$ Hz in the
DSSD detectors) and the coincidence time window was sharp ($\approx
10$ns), the effect of random coincidences was negligible. In the
following, we show only the real-coincidence gated spectra.

\begin{figure}[htb]
\begin{center}
{\includegraphics[scale=0.40]{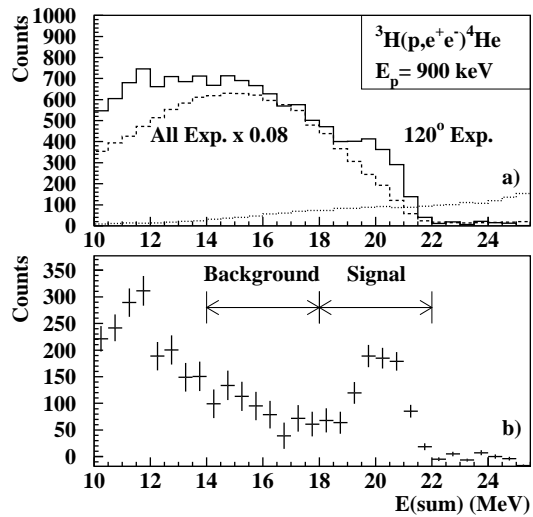}}\hspace{0.5cm}          
\caption{\it a): Experimental  energy-sum spectra of the
  $e^+e^-$ pairs derived, respectively, for ``All'' different detector
  combinations (dashed-line histogram with counts multiplied by 0.08)          
  and for detectors at 120$^\circ$ relative angles (solid-line histogram).      
  The cosmic-ray background contributions are subtracted from       
  both spectra. The CRB spectrum corresponding to the  ``All''      
  spectrum is plotted with dotted line.                             
 b): The difference of the solid-line and the dashed-line histograms above.}   
\end{center}
\vspace{-0.5cm}
\end{figure}
 
The downscaled (x0.08) energy-sum spectrum of the $e^+e^-$ pairs collected
by all combinations of the telescope pairs in the 10 - 25
MeV energy range is shown by a dashed-line histogram in Fig. 1a) after 
subtracting the cosmic-ray background (CRB). This
background was measured for two weeks before and after the experiments
using the same gates and conditions as used for the in-beam data. It has
been found that above E(sum)=25~MeV only the CRB 
contributes to the spectrum. 

The CRB contribution in the studied 10 - 25 MeV energy range has been 
determined by normalizing the off-beam spectrum to the in-beam spectra 
in the energy range E(sum)$\geq$25 MeV.
 This
contribution has been found to be a relatively small part of the total
spectrum as indicated by the dotted histogram in Figure 1a). In the figure,
 E(sum) means the measured energy sum of the
$e^+$ and $e^-$ particles corrected for the energy loss due to the
pair creation (1.02 MeV) and for the average energy loss of the two
particles when crossing the vacuum chamber and the DSSD detectors
(1.08 MeV).

In order to reduce the External Pair Creation (EPC) background, we constructed a
spectrum also from $e^+e^-$ pairs, which were detected by telescope
pairs with relative angles of 120$^\circ$. The spectrum is shown
in Fig. 1a) as a full-line histogram. The general shape of this
histogram is similar to that of the dashed-line one, however there is a well
observable peak on top of the smoothly decreasing shape at around 20.5
MeV. The difference of the two spectra is presented in Fig. 1b). 
The peak in the spectrum may come from the internal pairs created in the direct
proton capture process or in the 0$^+ \rightarrow 0^+$ E0 transition
of $^4$He and may also come from the $e^+e^-$-decay of the X17 hypothetical
particle.

To check these possibilities, in the
further analysis we have compared the angular correlation spectra of
the $e^+e^-$ pairs corresponding to this peak region (``Signal'' in Fig.1b)
with that of obtained for the  background region
(``Background'' in Fig. 1b).

The angular correlations of the $e^+e^-$  pairs were determined from the
position data of the DSSD detectors for each beam energy. First, the
efficiency calibration of the spectrometer was performed by taking the
uncorrelated pairs of consecutive events of the in-beam data as
described in detail in Ref. \cite{kr16}. This efficiency was then used to
normalize the raw angular correlation spectra, and finaly, the CRB
contributions were subtracted. Considering the kinematics of the
$e^+e^-$  pair creation process, we also required the following condition
for the asymmetry parameter: $-0.3\leq y=(E_{e^+} - E_{e^-})/(E_{e^+} + E_{e^-}) \leq 0.3$, where
$E_{e^+}$ and $E_{e^-}$ denote the kinetic energies of the positron
and electron, respectively.  The resulting angular correlation
spectra are indicated in Fig. 2 by dots, stars and full circles for
Ep= 510 keV, 610 keV and 900 keV, respectively. For better
readability, the spectra are shifted by 1-1 order of magnitude
according to the labels.

\begin{figure}[htb]
    \begin{center}
        {\includegraphics[scale=0.40]{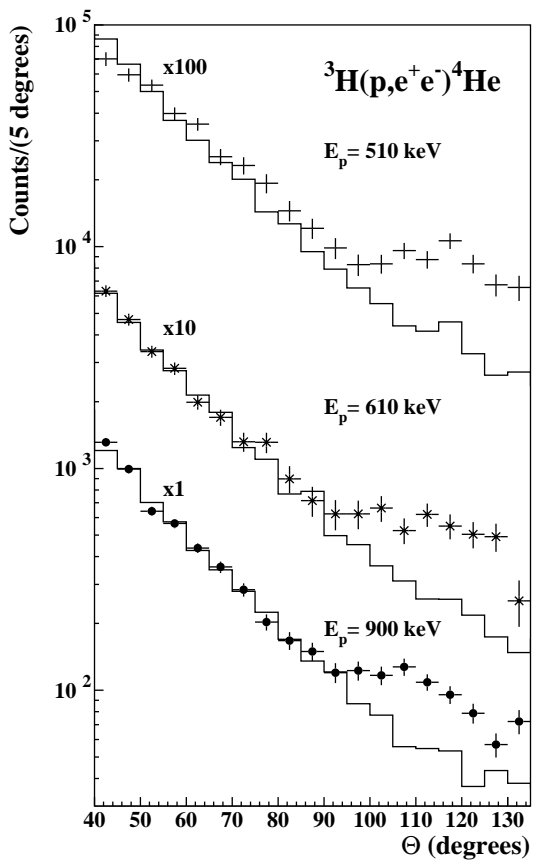}}\hspace{0.5cm}          
\caption{\it
Angular correlations of the $e^+e^-$ pairs for the "Signal" region
(see Fig. 1). Symbols with error bars indicate experimental data
measured in the $^{3}$H(p,$\gamma$)$^{4}$He reaction at different
proton beam energies, while solid-line histograms correspond to the
respective data obtained in the simulations described in the text.}
    \end{center}
\vspace{-0.5cm}
\end{figure}

The angular correlations of the $e^+e^-$ pairs for the background region marked in Fig. 1
are shown in Fig. 3 similarly to Fig. 2.

\begin{figure}[ht]
    \begin{center}
        {\includegraphics[scale=0.40]{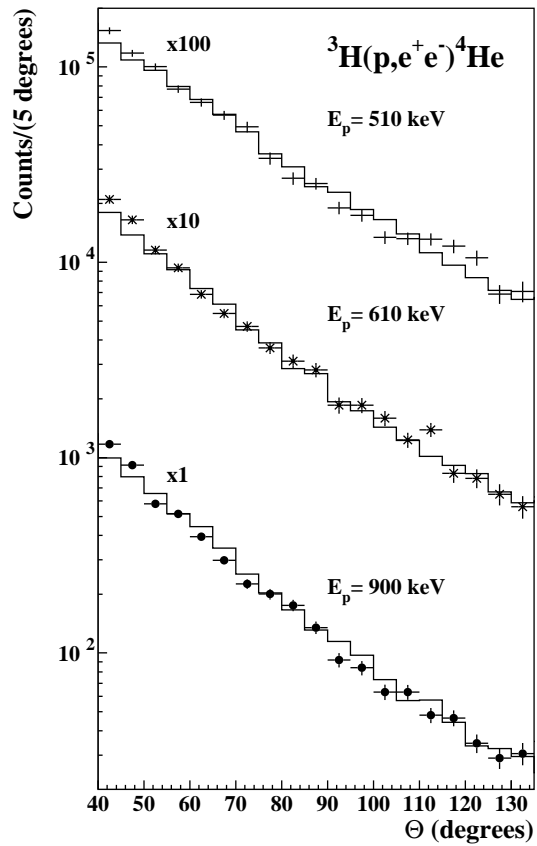}}\hspace{0.5cm}          
\caption{\it
Angular correlations of the $e^+e^-$ pairs for the  "Background" region (see Fig. 1). 
See the caption of Fig. 2 for more details.}
    \end{center}
\vspace{-0.5cm}
\end{figure}

The experimental angular correlations were compared to combinations of
Monte Carlo simulations of different processes resulting valid
$e^+e^-$ events in the spectrometer. An $e^+e^-$ event is considered valid,
if it passed all the conditions and cuts that was applied for the
experimental data. The simulations were performed by the
Geant3 code implementing the full experimental setup, the target, its
backing and the target holder. The energy resolution of the telescopes
was also considered and the energy sum of the simulated $e^+e^-$ pairs
should be either in the “Signal” or in the "Background" energy
range. Each of the simulated angular correlations was calculated by
taking into account the full stopping of the protons in the target.

In the light of the above considerations, we simulated the
contribution of the external $e^+e^-$ pairs created by the high-energy
$\gamma$-rays from the $^3$H(p,$\gamma$)$^4$He reaction. We also
determined the contribution of internal pair creation (IPC)
process. For that, both the properties of the proton capture process
and the emission of an $e^+e^-$ pair mediated by a one-photon exchange
was calculated by Viviani and co-workers \cite{vi21}. They provided us
high-statistic Monte-Carlo event files which we used as particle
generator inputs in our Geant3 simulations.

The simulated angular correlations are indicated by full-line histograms 
in Fig. 2. and Fig. 3.  for the "Signal" and the "Background"
sum-energy range, respectively.

We found that the most significant background was provided by $e^+e^-$ pairs 
created by $\gamma$-rays generated during direct proton capture on $^3$H.
For small correlation
angles, this process can fully interpret the measured values. 
As a result of this, the normalization of the contribution of $\gamma$ events 
was derived from the simulation's fit to the data in the 45 to 70  
degree opening angle region.

Note, that for the background regions of the sum-energy, the
experimental and the corresponding simulated curves (performed also for 
the background region)  showed a fairly good agreement over the
entire angular range (see Fig. 3), thus validating the correctness of
the simulations.

Here we mention that in this case the usual method of background
determination, i.e. performing the experiment without target material,
cannot be applied because the main source of the background is the
target material itself.

For further theoretical interpretation of the results shown in Fig. 2, 
the simulated angular correlations were subtracted from the experimental ones. 
The angular correlations (points with error bars) obtained after subtraction are
shown in Fig 4. 

\begin{figure}[htb]
    \begin{center}
        {\includegraphics[scale=0.45]{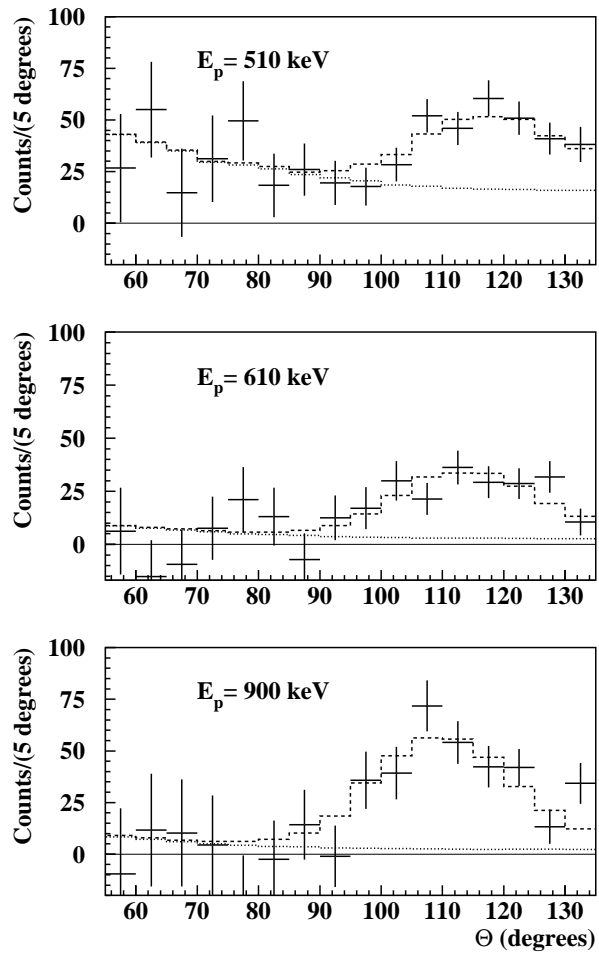}}\hspace{0.5cm}          
\caption{\it 
Comparison of the experimental and the simulated angular correlations of the        
$e^+e^-$ pairs. The best fitted sum (dashed line) as sum of the the simulated background   
(dotted line) and the simulated contribution of the hypothetical X17 boson is      
compared with the experimental signal values (dots with error bars).}              
    \end{center}
\vspace{-0.5cm}
\end{figure}

The corresponding proton beam energies are indicated in the figure. The
anomaly previously observed and explained by the decay of the X17
particle appeared at each of the bombarding energies.

In order to derive the exact value for the mass of the decaying
particle from the present data, we carried out a fitting procedure for
both the mass value and the amplitude of the observed peak.

The fit of the original experimental data was performed with 
RooFit \cite{Verkerke:2003ir} by describing
the $e^+e^-$ angular correlation with the following intensity function
($INT$):
\begin{equation}
\begin{aligned}
&INT = N_{EPC} * PDF(EPC) + N_{IPC} * PDF(IPC) + \\
&N_{Sig} * PDF(sig)\ ,
\end{aligned}
\label{eq:pdf}
\end{equation}

\noindent
where $PDF(X)$ stands for the MC-simulated probability density function 
and $N_X$ is the fitted number of the events of the given process. 
$PDF(sig)$ was simulated by Geant3 incorporating the relativistic two-body decay 
of a particle with a given mass. Therefore, $PDF(sig)$ was constructed
as a 2-dimensional model function of
the $e^+e^-$ opening angle and the mass of the simulated particle. To
construct the mass dependence, the PDF linearly interpolates the
$e^+e^-$ opening angle distributions simulated for discrete particle
masses.

Using the intensity function described in Equation~\ref{eq:pdf}, we first
performed a list of fits by fixing the simulated particle mass in the
signal PDF to a certain value, and employing RooFit estimate the best
values for $N_{Sig}$ and $N_{EPC}$ and $N_{IPC}$. 
Allowing the particle mass to vary 
in the fit, the best fitted mass is calculated.
We also made fits by fixing the $N_{EPC}$ value to the experimental data in the
angular range of 40 to 70 degrees.

The values obtained for the energy and branching ratio of the X17
boson and the IPC values, as a result of the average of the two fits 
described above,  are summarized in Table 1, and the corresponding fits  
after subtraction of the EPC background are shown in Fig. 4.

\begin{table}[!h]
\centering
\caption{Internal Pair Creation Coefficients (IPCC), X17 Boson
  branching ratios (Bx), masses of the X17 particle,  and confidences
 derived from the fits.}
\label{tab-1}       
\begin{tabular}{llllc}
\hline\hline
E$_p$ & IPCC          & B$_x$           &Mass& Confidence \\\hline
 (keV)&$\times10^{-4}$&$\times10^{-6}$&(MeV/c$^2$)\\\hline
510 & 2.5(3)   & 6.2(7) & 17.01(12)& 7.3$\sigma$\\
610 & 1.0(7)   & 4.1(6)  & 16.88(16)& 6.6$\sigma$\\
900 & 1.1(11)   & 6.5(20) & 16.68(30)& 8.9$\sigma$\\\hline
Averages& & 5.1(13) & 16.94(12)&            \\\hline             
$^8$Be values& & 6 & 16.70(35)&      \\\hline\hline
\end{tabular}
\end{table}

Table 1. shows the fitting parameters and the average of the
parameters. As can be seen,  consistent values were obtained for each
fitting parameter.  In the last row, our corresponding values measured
in the case of $^8$Be are also shown \cite{kr16}.

As shown, the branching ratios of the X17 particle are identical within
error bars, for the three beam energies proving that the X17
particle was most likely formed in direct proton capture, which has a
dominant multipolarity of E1.

As discussed, IPC generated mostly during the direct capture (E1) transition, 
however, the IPCC deduced to be much smaller than expected from the Bohr's 
approximation and also smaller than the one predicted by Viviani and 
co-workers \cite{vi21}.

The systematic uncertainties were estimated by simulations. Taking into account the
uncertainty of the target position along the beam line
estimated to be $\pm$ 2 mm, may cause an $\Delta
m_\mathrm{X}c^2$=$\pm$ 0.06 MeV uncertainty. 
The uncertainty of the
place of the beam spot perpendicular to the beam axis was estimated to
be $\pm$ 2 mm in the worst case, which may cause a shift in the
invariant mass of $\Delta m_\mathrm{X}c^2$$\pm$=0.15 MeV. The
whole systematic error was conservatively estimated as: $\Delta
m_\mathrm{X}c^2$(syst.) = $\pm$0.21 MeV.

We emphasize, that the obtained mass agrees very well with that observed in the earlier
$^8$Be experiment ($m_\mathrm{X}c^2$=16.70$\pm 0.35 (stat.) \pm 0.5
(syst.)$ MeV), which is remarkable considering that the excesses in
the observed angular correlation spectra appear at different
correlation angles as one would indeed expect from the kinematics of 
the relativistic two-body decay. 
Thus, we strongly believe, that our new observation enhance the possibility 
that the measured anomalies can be attributed to the same new particle X17.

Very recently, Zhang and Miller \cite{zh21} studied the protophobic
vector boson explanation by deriving an isospin relation between
photon and X17 couplings to nucleons. They concluded that X17
production is dominated by direct capture transitions both in $^8$Be
and $^4$He without going through any nuclear resonance.  A smooth
energy dependence is predicted that occurs for all proton beam
energies above threshold \cite{zh21}.  Our present results obtained
for $^4$He at different beam energies agrees with their prediction.

In summary, we have studied the energy-sum and angular correlation
spectra of $e^+e^-$ pairs produced in the $^{3}$H(p,$\gamma$)$^{4}$He
reaction at $E_p$= 510, 610 and 900 keV proton energies. 
The main features of the
spectra can be understood rather well taking into account the internal
and external pair creations following the direct proton capture on the
target.  We have, however, observed a peak-like anomalous
excess of $e^+e^-$ pairs in the angular correlation spectra around
115$^\circ$ at each beam energy. 
This $e^+e^-$ excess cannot be accounted for by the above
processes, however, it can be described by the creation and subsequent
decay of a light particle during the proton capture process to the ground
state of the $^{4}$He nucleus. The derived mass of the particle
($m_\mathrm{X}c^2$=16.94$\pm 0.12 (stat.) \pm 0.21 (syst.)$~MeV) agrees         
well with that of the X17 particle, which we recently suggested
\cite{kr16,kra17,kra19} for describing the anomaly observed in $^8$Be.
This observation supports the X17 particle hypothesis.


\begin{acknowledgements}
We wish to thank M. Viviani for providing us their theoretical results,
D. Horv\'ath for the critical reading
of the manuscript and for the many useful discussions. We wish to thank
also for Gy. Zilizi for providing us the $^3$H targets and  for Z. Pintye
for the mechanical and J. Moln\'ar for the electronic design of the experiment.     
This work has been supported by the Hungarian NKFI Foundation No.\,
K124810 and by the GINOP-2.3.3-15-2016-00034 and 
GINOP-2.3.3-15-2016-00034 grants.
\end{acknowledgements}

\end{document}